\documentclass[]{aa}  
\usepackage{natbib} 
\bibpunct{(}{)}{;}{a}{}{,} 
\usepackage{graphicx}
\usepackage{txfonts}
\def\inte{{\em INTEGRAL}}

\def\rxte{{\em RXTE}}
\def\swift{{\em Swift}}

\def \inte {{$INTEGRAL$}}

\def \ferg {erg~cm$^{-2}$~s$^{-1}$}
\def \hcm {\hbox {\ifmmode $ atom cm$^{-2}\else atom cm$^{-2}$\fi}}
\def \arcmin {\hbox{$^\prime$}}

\def \igr {IGR\,J19294+1816}

\begin{document}
   \title{\inte\ and \swift\ observations of \igr\ in outburst}

\author{E. Bozzo 
   \inst{1}
	\and 
   C. Ferrigno 
 	\inst{1} 
 	\and   
   M. Falanga  
   \inst{2}  
	\and	
	R. Walter 
		\inst{1}		
}

\authorrunning{E. Bozzo et al.}
  \titlerunning{ \igr\ in outburst }
  \offprints{E. Bozzo}

\institute{ISDC data center for astrophysics of the University of Geneva
	 chemin d'\'Ecogia, 16 1290 Versoix Switzerland. 
	\email{enrico.bozzo@unige.ch}
         \and
        International Space Science Institute (ISSI) Hallerstrasse 6, CH-3012 Bern, Switzerland.   
         }

 \abstract{ \igr\ was discovered by \inte\ in 2009 during a bright X-ray outburst and was classified as a possible 
 Be X-ray binary or supergiant fast X-ray transient.}
 {On 2010 October 28, the source displayed a second X-ray outburst and a two months-long 
 monitoring with \swift\ was carried out to follow the evolution of the source X-ray flux 
 during the event.}{We report on the \inte\ and \swift\ observations of the second X-ray outburst observed from 
 \igr.\ }{We detected pulsations in the X-ray emission from the source at $\sim$12.5~s up to 50~keV. The source X-ray flux 
 decreased smoothly during the two months of observation, displaying only marginal spectral changes. 
 Owing to the relatively rapid decay of the source X-ray flux, no significant variations of the source spin period across the event could be 
 measured. This prevented a firm confirmation of the previously suggested orbital period of the source at   
 117~d. This periodicity was also searched for in archival \swift\,/BAT data.  
 We detected a marginally significant peak in the periodogram and determined the best period at 
 116.2$\pm$0.6~days (estimated chance probability of a spurious detection 1\%).}
 {The smooth decline of the source X-ray flux across the two months of observations after 
 the onset of the second outburst, together with its relatively low value of the spin period and the absence of 
 remarkable changes in the spectral parameters (i.e., the absorption column density), suggests that \igr\ is most 
 likely another member of the Be X-ray binaries  discovered by \inte\ and not a supergiant fast X-ray transient.}


   \date{Received 2011 Feb. 15; accepted 2011 May 07}

   \maketitle

\section{Introduction}
\label{sec:intro} 

\igr\ was discovered undergoing 
a bright X-ray outburst for the first time by \inte\ on 2009 March 27 
\citep{turler09}.  
On this occasion, \igr\ was detected by the IBIS/ISGRI telescope   
at a significance level of 8.3~$\sigma$ in the 20-40 keV energy band. 
The source was not detected in the 40-80~keV energy band.  
The IBIS/ISGRI spectrum (18-50 ~keV) could be reasonably well fitted with 
a power-law model. The estimated photon index was $\Gamma$=4.0$\pm$0.7 
and the corresponding flux 14$\pm$2 mCrab (i.e., 1.1$\times$10$^{-10}$~\ferg). 

The analysis of archival \swift\ observations permitted obtaining  
a refined position for the source at $\alpha_{\rm J2000}$=19$^{\rm h}$29$^{\rm m}$55$\fs$9 and 
$\delta_{\rm J2000}$=18${\degr}$18$\arcmin$38$\farcs$4 (associated 
uncertainty of 3.5$\farcs$ at 90\% c.l.), and thus identify the  
IR counterpart to \igr\ as the object 2MASS\,J19295591+1818382 
\citep{rodriguez09}. These data also showed evidence of pulsations 
at $\sim$12.5~s, later confirmed with \rxte\ \citep{strohmayer09}. 

The historical \swift\,/BAT lightcurve of the source 
revealed the presence of a modulation with a period of 117~d, 
which was tentatively interpreted as the orbital period of the system 
\citep{krimm09, corbet09}. 

A summary of all these results and an in-depth 
inspection of all available \inte\ observations performed in the 
direction of the source was provided by \citet{rodriguez09b}.   
The authors reported on the discovery of relatively short 
flares (few thousands of seconds) from this source with \inte,\ and 
concluded that, even though the orbital and spin period of 
\igr\ would nicely place the source in the region of the $P_{\rm spin}$-$P_{\rm orb}$ 
diagram populated by the Be X-ray binaries, a possible supergiant fast X-ray 
\citep[SFXTs, see e.g.,][]{walter07} nature of this source could not be ruled out. 

\igr\ was detected undergoing a renewed activity with \inte\ on 
2010 October 28 \citep{bozzo10,jenke10}. Motivated by the previous 
findings, we triggered a monitoring program of this source 
with \swift\,/XRT. This permitted us  for the first time to follow in detail 
the evolution of the source X-ray flux for about two months 
after the new outburst. 
In this paper, we report on the results of the \inte\,/ISGRI and \swift\,/XRT observations 
carried out during the outburst in 2010 and reanalyze all previously available 
observations performed with these instruments to investigate the real nature of the source. 
In Sect.~\ref{sec:data} we describe the data analysis technique and the results. 
Our discussion and conclusions are summarized in Sect.~\ref{sec:discussion}.

\section{Data analysis}
\label{sec:data}

\subsection{ \inte\ }
\label{sec:integral}

\inte\ observations are commonly divided into ``science windows'' (SCWs), 
i.e., pointings with typical durations of $\sim$2-3~ks.  
We considered all available SCWs for the IBIS/ISGRI 
\citep[17-80~keV,][]{lebrun03,ubertini03} telescope that were performed in the 
direction of \igr\ close to the period in which a renewed activity from this 
source was discovered (see Sect.~\ref{sec:intro}). 
These comprised observations of the region around GRS1915+105 in satellite revolutions 980, 
982, 985, 986, and 988.  
A complete log of the observations is provided in Table~\ref{tab:integral}. 
The source was always outside the field of view (FOV) of the lower energy telescope 
JEM-X on-board \inte\ \citep[][]{lund03}. 
All \inte\ data were analyzed using version 9.0 of the OSA 
software distributed by the ISDC \citep{courvoisier03}. 
\begin{figure}
\centering
\includegraphics[height=6.5cm,width=9cm]{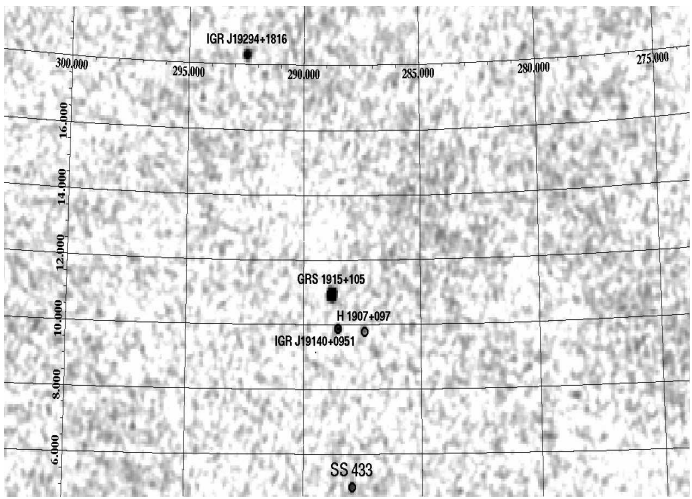}
\caption{Mosaic of the IBIS/ISGRI FOV around \igr\ realized by using all available SCWs  
in the direction of the source during satellite revolution 982 (20-50~keV energy range).}  
\label{fig:integral} 
\end{figure} 

\begin{figure}
\centering
\includegraphics[scale=0.35,angle=-90]{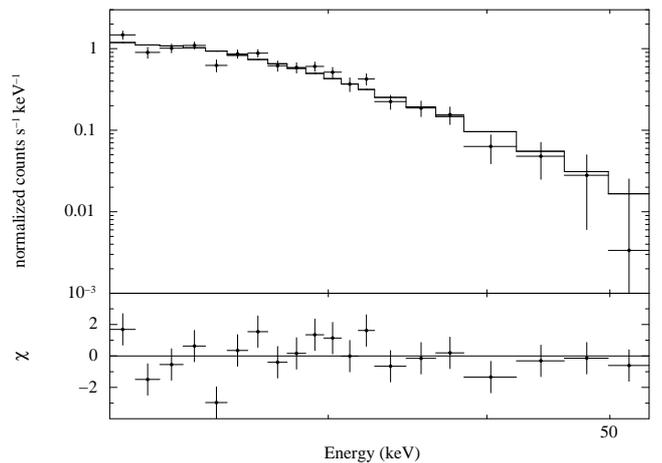}
\caption{IBIS/ISGRI spectrum of \igr\ extracted from the observations in 
Rev.~982. The best fit is obtained with a simple power-law model and is shown 
in the figure with a continuous line. The residuals from this fit are reported in the 
bottom panel.}   
\label{fig:spectrum} 
\end{figure}

\igr\ was detected in rev.~982 at a significance level of 27.3~$\sigma$ in the 20-50~keV 
energy band \citep[this energy band was chosen to maximize the S/N, see][]
{bozzo10}. An image of the IBIS/ISGRI FOV around \igr\ is shown in 
Fig.~\ref{fig:integral}. Given the relatively high detection significance, we rebinned the 
IBIS/ISGRI response matrix\footnote{See http://isdcul3.unige.ch/Soft/download/osa/osa\_doc/ 
osa\_doc-9.0/osa\_um\_ibis-9.2/.} to extract a spectrum with 
22 energy bins and look for possible spectral features. A reasonably good fit to this spectrum 
was obtained using a simple power-law model ($\chi^2_{\rm red}$/d.o.f.=1.4/18). We measured a power-law photon index of 
$\Gamma$=4.9$\pm$0.3 and estimated a source flux of (5.4$\pm$0.3)$\times$10$^{-10}$~erg/cm$^2$/s 
(20-50~keV).  
The IBIS/ISGRI spectrum of \igr\ is shown in Fig.~\ref{fig:spectrum} together with the 
best-fit model. No convincing spectral features emerged from the residuals of this fit. 

We also analyzed the closest \inte\ observations performed in the direction of the source 
before the discovery of its renewed activity. \igr\ was in the IBIS/ISGRI FOV 
from 2010 October 22 at 23:04 to 2010 October 23 at 02:21 (UTC; satellite revolution 980). 
In this case the source was not detected. We estimated a 3~$\sigma$ upper limit on its X-ray flux of 5.1~mCrab 
(20-50~keV energy band; effective exposure time 22~ks). 
This corresponds\footnote{The conversion between the source count-rate 
and mCrab is carried out by using the IBIS/ISGRI observation of the Crab in 
Rev.~967. Here the count-rate of the Crab in the 20-50~keV energy band was 168.86$\pm$0.09 
and the corresponding flux 1.0$\times$10$^{-8}$~\ferg} to 5$\times$10$^{-11}$~\ferg.   
\begin{figure}
\centering
\includegraphics[scale=0.35,angle=-90]{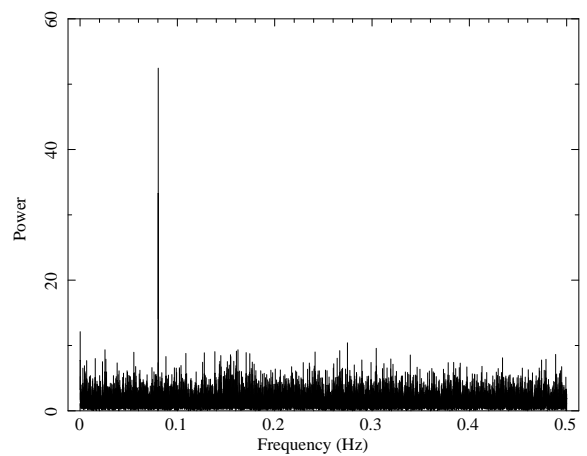}
\caption{Power spectrum of the IBIS/ISGRI data of \igr\ in Rev.~982. Pulsations at a frequency of 
0.080312(6)~Hz (error at 90\% c.l. on the last digit) are clearly detected.}
\label{fig:powerspectrum} 
\end{figure}

A similar analysis was also performed on the IBIS/ISGRI mosaic realized by using the available 
\inte\ observations including \igr\ in their FOV carried out immediately after 
the discovery of the renewed activity (satellite Rev.~985, 986, 988). 
During this period (from 2010 November 6 at 17:46 to 2010 November 18 at 00:26 UTC) 
the source was not detected and we estimated a 3~$\sigma$ upper limit on its X-ray 
flux in the 20-50~keV energy band of 3.1~mCrab (i.e. roughly 3$\times$10$^{-11}$~\ferg; 
effective exposure time 56~ks). 
\begin{figure}
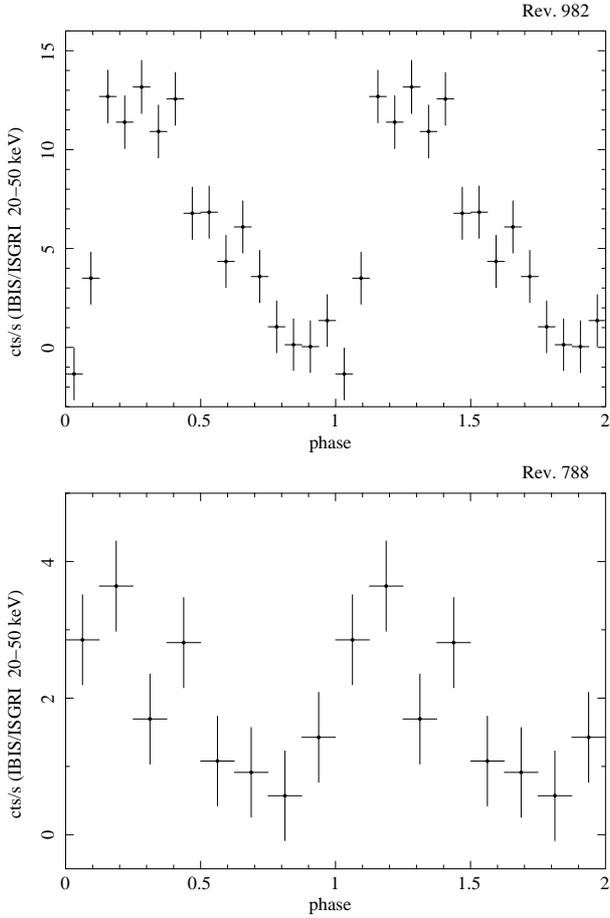

\centering
\includegraphics[scale=0.37,angle=-90]{pulseprofile982.ps}
\includegraphics[scale=0.37,angle=-90]{pulseprofile788.ps}
\caption{IBIS/ISGRI lightcurve of \igr\ in rev.~982 (top panel) and 
rev.~788 (bottom panel) folded at the best determined spin frequency. 
In both cases we used data in the 20-50~keV energy range.}
\label{fig:pulseprofile} 
\end{figure}

Pulsations were searched in the \inte\ data of rev.~982 by extracting 
source and background lightcurves with time resolutions of 1~s applying the method 
described in \citet[][all event times were also 
corrected to the solar system barycenter]{segreto07}. Pulsations were 
clearly detected in these data (see Fig.~\ref{fig:powerspectrum}). 
We estimated the best pulse frequency at 0.080312(6)~Hz (error at 90\% c.l on the last digit) 
using standard phase-shift techniques \citep{ferrigno07}, 
and measured an rms fractional amplitude of 0.84$\pm$0.06 \citep[defined according 
to Eq.~14 in][]{klis88}. 
The IBIS/ISGRI pulsed profile of \igr\ is shown in Fig.~\ref{fig:pulseprofile}. 
\begin{table}
\tiny
\caption{ \inte\ observations log of \igr.\ } 
\begin{tabular}{@{}lllllllll@{}}
\hline
\hline
\noalign{\smallskip}
Rev.  & START TIME & STOP TIME & EXP      & $F_{\rm X}$ \\
      &    (MJD)   &   (MJD)   & (ks)     & (erg~cm$^{-2}$~s$^{-1}$) \\
\noalign{\smallskip}
\hline
\noalign{\smallskip}
788       & 54917.49852 & 54917.73836 & 12 & 1.1$\times$10$^{-10}$ \\
\noalign{\smallskip}
980$^{a}$ & 55491.96168 & 55492.22987 & 22 & $<$5.0$\times$10$^{-11}$ \\
\noalign{\smallskip}
982 & 55497.76780 & 55498.00752 & 3.4 & 5.4$\times$10$^{-10}$ \\
\noalign{\smallskip}
985,6,8$^{a}$ & 55506.74066 & 55518.01843 & 56 & $<$3.0$\times$10$^{-11}$ \\
\noalign{\smallskip}
\hline
\noalign{\smallskip}
\multicolumn{5}{l}{NOTE: $F_{\rm X}$ is the flux in the 20-50~keV energy band. EXP indicates the}\\  
\multicolumn{5}{l}{effective exposure time of each observation. $^{a}$: 3$\sigma$ upper limit.} \\ 
\end{tabular}
\label{tab:integral}
\end{table} 

In order to better compare these results with the previous findings, 
we also reanalyzed the \inte\ data of the previous outburst of 
\igr\ (Rev. 788, see Sect.~\ref{sec:intro}). 
In this case, we found no clear indication of pulsations in the IBIS/ISGRI 
periodogram (the source X-ray flux was a factor of $\sim$5 lower than that in rev. 982). 
However, an epoch-folding search around the previously know period 
led to a marginal detection of pulsations. We determined a best 
pulse frequency of 0.08036(2)~Hz, and estimated a 
fractional rms of 0.55$\pm$0.12. The IBIS/ISGRI 
lightcurve of this observation folded at the best determined 
spin frequency is shown in Fig.~\ref{fig:pulseprofile}. 

A log of all the \inte\ observations used 
is reported in Table~\ref{tab:integral}.

\subsection{ \swift\ }
\label{sec:swift}

Following the detection of a new outburst from \igr,\ 
we requested a monitoring observation of the source 
to follow closely the evolution of its X-ray 
flux down to the quiescent level. 
A log of all \swift\,/XRT observations is 
provided in Table~\ref{tab:swift}. 

\swift\,/XRT data were analyzed by using standard procedures 
\citep{burrows05} and the latest calibration files available. 
The XRT data were processed with the {\sc xrtpipeline} 
(v.0.12.6); filtering and screening criteria were applied 
by using {\sc ftools} ({\sc Heasoft} v.6.10). 
We extracted source and background light 
curves and spectra by selecting event grades of 0-2 and 0-12   
for the window time and photon counting (PC) mode, respectively. 
Exposure maps were created through the {\sc xrtexpomap} task, and we used 
the latest spectral redistribution matrices in the {\sc heasarc} 
calibration database (v.011). 
Ancillary response files, accounting for different 
extraction regions, vignetting and PSF corrections, were generated 
by using  the {\sc xrtmkarf} task.  
When required, we corrected PC data for pile-up, and used the 
{\sc xrtlccorr} to account for this correction in the 
background-subtracted light curves. Source and background event lists 
were extracted by using the highest possible time resolution for the 
PC (2.5~s) and the WT (0.0018~s) observational modes, and 
then barycentered by using the {\sc barycorr} tool available 
within the {\sc Heasoft} software package.  
\begin{table*}
\tiny
\caption{ \swift\ observations log of \igr.\ } 
\begin{tabular}{@{}lllllllll@{}}
\hline
\hline
\noalign{\smallskip}
OBS ID & INSTR & START TIME & STOP TIME & EXP & $N_{\rm H}$ & $\Gamma$ & $F_{\rm obs}$ & $\chi^2_{\rm red}$/d.o.f. \\
       &       &     (MJD)       &   (MJD)        &  (sec)   &  (10$^{22}$~cm$^{-2}$) &  & (erg~cm$^{-2}$~s$^{-1}$) & (C-stat/d.o.f.) \\
\noalign{\smallskip}
\hline
\noalign{\smallskip}
{\scriptsize {\bf 2007}} \\ 
\noalign{\smallskip}
\hline
\noalign{\smallskip}
00037191001 & XRT/PC & 54443.05208  & 54444.94375  & 7.9 & 4.4$^{+1.1}_{-0.8}$  & 0.8$^{+0.3}_{-0.2}$ & 3.2$\times$10$^{-11}$ & 0.9/70 \\
\noalign{\smallskip}
00037191002 & XRT/PC & 54447.01389  & 54447.35347  & 3.4 & 3.7$^{+1.3}_{-1.1}$  & 0.9$\pm$0.4         & 3.2$\times$10$^{-11}$ & 0.9/33 \\
\noalign{\smallskip}
\hline
\noalign{\smallskip}
{\scriptsize {\bf 2009}} \\ 
\noalign{\smallskip}
\hline
\noalign{\smallskip}
00031392001 & XRT/PC & 54925.80417  & 54926.00000  & 2.6 & 3.6$^{+6.3}_{-3.6}$  & 1.1$^{+1.5}_{-1.1}$ & 6.5$\times$10$^{-12}$ & (17.4/19) \\
\noalign{\smallskip}
00031392002$^{a}$ & XRT/PC & 54973.86319  & 54973.93889  & 2.0 & 3.6 (fixed)          & 1.1 (fixed)         & 2.0$\times$10$^{-12}$ & --- \\
\noalign{\smallskip}
00031392003$^{a}$ & XRT/PC & 54974.53819  & 54974.67153  & 1.7 & 3.6 (fixed)          & 1.1 (fixed)         & 2.2$\times$10$^{-12}$ & --- \\
\noalign{\smallskip}
00031392004$^{b}$ & XRT/PC & 54978.02153  & 54978.16319  & 1.2 & 3.6 (fixed)          & 1.1 (fixed)         & 1.3$\times$10$^{-12}$ & --- \\
\noalign{\smallskip}
\hline
\noalign{\smallskip}
{\scriptsize {\bf 2010}} \\ 
\noalign{\smallskip}
\hline
\noalign{\smallskip}
00031854001 & XRT/PC & 55498.62225  & 55498.63470  & 0.99 & 3.7$^{+0.8}_{-0.7}$  & 1.1$\pm$0.3         & 2.6$\times$10$^{-10}$   & 1.8/52 \\
\noalign{\smallskip}
00031854002 & XRT/WT & 55498.69269  & 55498.88016  & 0.12 & 4.4$^{+2.3}_{-1.6}$  & 1.5$^{+0.7}_{-0.6}$ & 2.6$\times$10$^{-10}$   & 0.8/11 \\
            & XRT/PC &                 &                 & 2.6  & 3.0$^{+0.6}_{-0.5}$  & 1.0$\pm$0.2         & 2.3$\times$10$^{-10}$   & 0.7/72 \\
\noalign{\smallskip}
00031854003 & XRT/PC & 55500.21293  & 55500.22519  & 0.95 & 10.7$^{+5.0}_{-4.0}$ & 0.7$\pm$0.7         & 6.2$\times$10$^{-11}$   & (46.8/50) \\
\noalign{\smallskip}
00031854004 & XRT/PC & 55501.63127  & 55501.64450  & 0.97 & 7.4$^{+3.9}_{-2.6}$  & 1.2$^{+0.7}_{-0.6}$ & 4.5$\times$10$^{-11}$   & (48.1/48) \\
\noalign{\smallskip}
00031854005 & XRT/PC & 55503.63889  & 55503.97292  & 1.3  & 7.0$^{+4.1}_{-3.4}$  & 1.0$^{+0.9}_{-0.7}$ & 3.0$\times$10$^{-11}$   & (34.8/41) \\
\noalign{\smallskip}
00031854006$^{a}$ & XRT/PC & 55505.50556  & 55505.71944  & 1.4  & 7.0 (fixed)          & 1.0 (fixed)         & 8.2$\times$10$^{-12}$   & ---  \\
\noalign{\smallskip}
00031854007 & XRT/PC & 55507.79097  & 55507.92986  & 1.0  & 4.8$^{+2.8}_{2.2}$   & 1.5$^{+0.8}_{-0.4}$ & 2.5$\times$10$^{-11}$   & (43.0/36) \\
\noalign{\smallskip}
00031854008 & XRT/PC & 55510.45347  & 55510.53889  & 2.4  & 4.7$^{+4.0}_{-3.6}$  & 1.1$^{+1.0}_{-0.9}$ & 1.5$\times$10$^{-11}$   & (17.0/27) \\
\noalign{\smallskip}
00031854009$^{a}$ & XRT/PC & 55513.46458  & 55513.54931  & 2.2  & 4.7 (fixed)          & 1.1 (fixed)         & 4.3$\times$10$^{-12}$    & --- \\
\noalign{\smallskip}
00031854010$^{a}$ & XRT/PC & 55516.56111  & 55516.63472  & 2.3 &  4.7 (fixed)          & 1.1 (fixed)         & 1.0$\times$10$^{-11}$    & --- \\
\noalign{\smallskip}
00031854011 & XRT/PC & 55523.03056  & 55523.10972  & 2.3 & 6.4$^{+4.4}_{-4.2}$   & 1.8$^{+1.2}_{-1.3}$ & 8.1$\times$10$^{-12}$   & (20.0/24) \\
\noalign{\smallskip}
00031854012 & XRT/PC & 55530.05625  & 55530.26944  & 2.8 & 4.8$^{+2.3}_{-2.1}$   & 1.2$\pm$0.7         & 9.3$\times$10$^{-12}$   & (28.0/31) \\
\noalign{\smallskip}
00031854013 & XRT/PC & 55537.01806  & 55537.10139  & 2.2 & 8.3$^{+1.7}_{-1.5}$  & 2.8$^{+2.1}_{-1.7}$  & 3.4$\times$10$^{-12}$   & (13.2/11) \\
\noalign{\smallskip}
00031854014 & XRT/PC & 55544.10764  & 55544.99028  & 3.3 & 4.4$^{+3.0}_{-3.2}$  & 1.3$^{+1.0}_{-1.1}$  & 6.9$\times$10$^{-12}$   & (19.4/24) \\
\noalign{\smallskip}
00031854015 & XRT/PC & 55550.47847  & 55550.62917  & 3.4 & 4.6$^{+2.9}_{-2.5}$  & 1.2$^{+0.9}_{-0.8}$  & 5.6$\times$10$^{-12}$   & (11.8/21) \\
\noalign{\smallskip}
\hline
\noalign{\smallskip}
\multicolumn{9}{l}{NOTE: Spectra extracted from these observations are fitted with an absorbed power law 
(absorption column density N$_{\rm H}$ and photon index $\Gamma$).} \\
\multicolumn{9}{l}{$F_{\rm obs}$ is the absorbed flux in the 1-10~keV energy band. 
EXP indicates the total exposure time of each observation} \\ 
\multicolumn{9}{l}{$^{a}$: We used {\sc sosta} and {\sc webpimms} to estimate this flux. $^{b}$: 90\% c.l. upper limit.}\\
\end{tabular}
\label{tab:swift}
\end{table*} 

\begin{figure}
\centering
\includegraphics[scale=0.5]{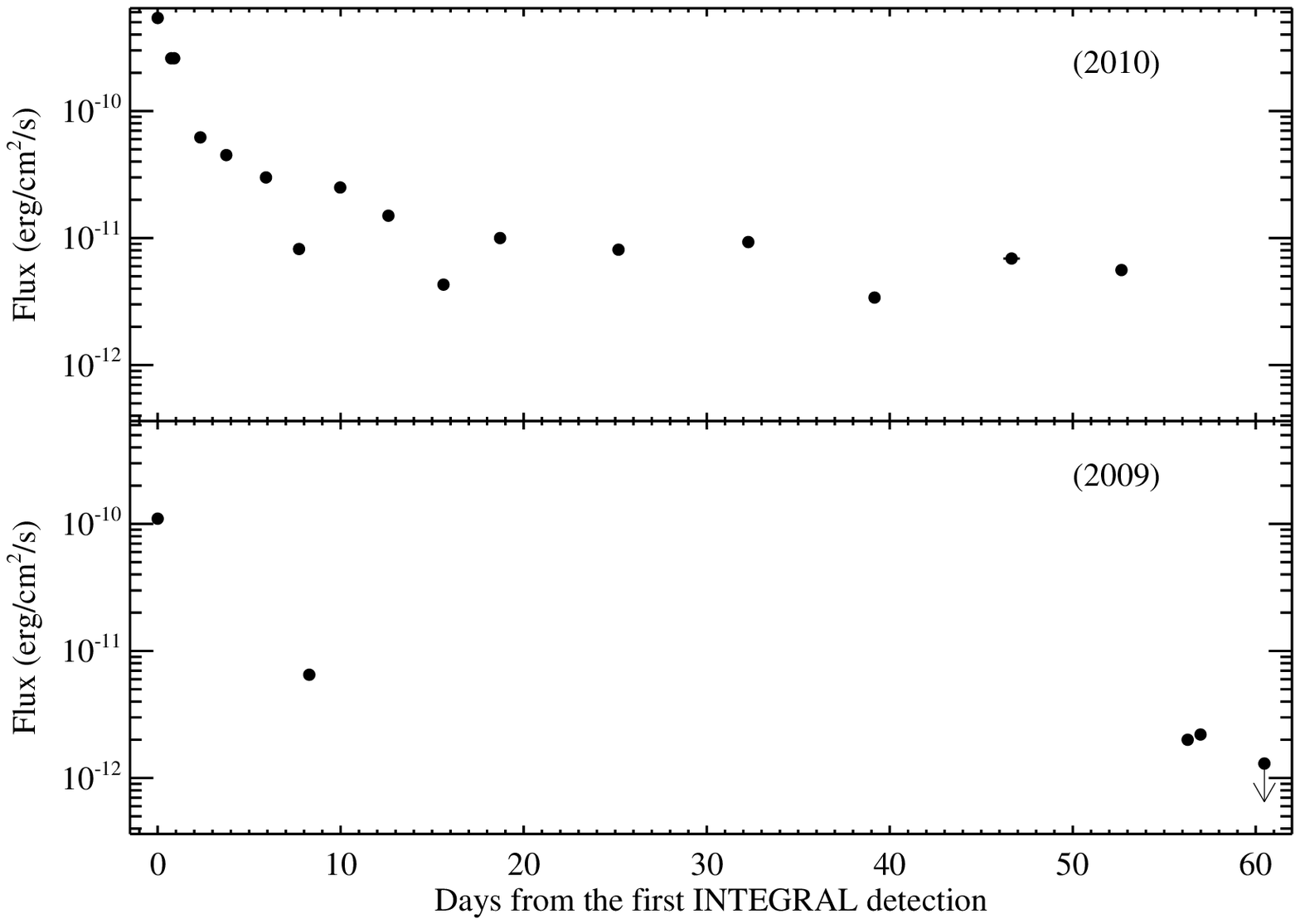}
\caption{Lightcurves of the outburst decay of \igr\ during the two events that 
occurred in 2009 (lower panel) and 2010 (upper panel). We did not plot the observations carried out in 2007 
because in that case the beginning of the outburst was not caught by any X-ray satellite  
(see Tables~\ref{tab:integral} and \ref{tab:swift}). We scaled the values on the 
X-axis to the first \inte\ detection of the source (first point on the left in both panels). 
This corresponds to 55497.8877 MJD for the upper panel and 54917.6184 MJD for the lower panel.} 
\label{fig:lightcurvestot} 
\end{figure}

From the observations in Table~\ref{tab:swift} that were characterized by adequate 
statistics, we extracted source and background spectra and rebinned these spectra 
to have at least 20 photons per bin to permit $\chi^2$ fitting. 
We derived a mean X-ray flux by fitting these spectra with 
an absorbed power law model (we used {\sc Xspec} v.12.6.0). 
The spectra extracted during observations characterized by a lower statistic 
were instead rebinned to have at least five photons per bin and were then fitted by using the C-statistic.    
For the observations in which less than 100~counts were recorded from the source, we estimated the source 
count-rate with {\sc sosta} ({\sc ximage} V.4.5.1)  
and then used this count-rate within {\sc webpimms}\footnote{http://heasarc.nasa.gov/Tools/w3pimms.html}  
to derive the X-ray flux \citep[we assumed the same spectral model 
of the closest observation for which a proper spectral analysis could be carried out; see e.g.,][]{bozzo09}.  
All these results are reported in Table~\ref{tab:swift} and used in Fig.~\ref{fig:lightcurvestot} 
to plot the variation of the source X-ray flux during the decay from the outburst. 

\begin{figure}
\centering
\includegraphics[scale=0.35,angle=-90]{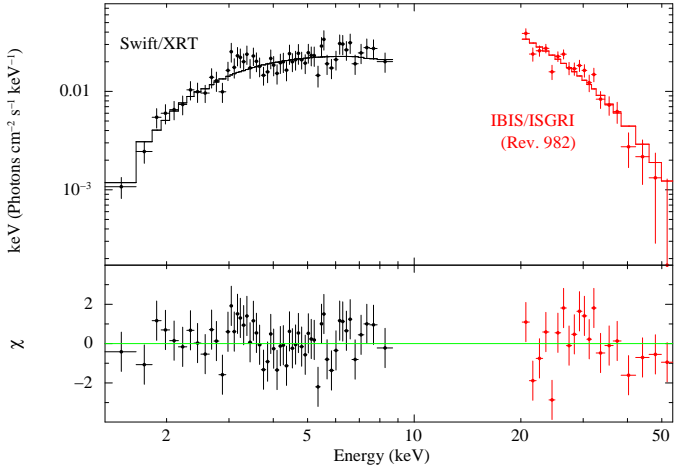}
\caption{Unfolded joint \swift\,/XRT and \inte\,/ISGRI spectrum (we used the 
XRT spectrum extracted from observation ID.~00031854001 and the ISGRI spectrum 
extracted from the \inte\ rev.~982). 
The best-fit model is obtained with a CUTOFFPL model (see text for details). 
The residuals from this fit are shown in the bottom panel.}
\label{fig:unfolded} 
\end{figure}

Even though the observations were not strictly simultaneous, we performed a broad-band 
(0.3-50 keV) fit to the source spectrum by combining the IBIS/ISGRI data from rev.~982 and the \swift\,/XRT data from 
the observation ID.~00031854001. We used first a simple absorbed blackbody or power-law model, and 
introduced a normalization constant to take into account both the intercalibration 
between the \swift\,/XRT and IBIS/ISGRI instruments and the variability of the source. 
Both fits gave unacceptable results with $\chi^2_{\rm red}$$>$2.3. We therefore substituted the power-law 
component with a cut-off power law ({\sc wabs*cutoffpl} in {\sc xspec}). This model gave an acceptable 
fit ($\chi^2_{\rm red}$/d.o.f.=1.1/71) and we measured a power-law photon index of $\Gamma$=0.4$\pm$0.3, 
an absorption column density of $N_{\rm H}$=(3.1$\pm$0.7)$\times$10$^{22}$~cm$^{-2}$ and a cut-off energy of 
8.0$^{+1.2}_{-1.0}$~keV. The normalization constant turned out to be 4.2$^{+1.7}_{-1.2}$, which indicates (as expected) 
that the flux of the source during the \swift\ observation already decreased with respect to that measured by \inte\ 
during satellite rev.~982 (see also Tab.~\ref{tab:integral} and \ref{tab:swift}). The unfolded XRT+ISGRI spectrum,  
the best-fit model and the residuals from this fit are shown in Fig.~\ref{fig:unfolded}. 

\begin{figure}
\centering
\includegraphics[scale=0.37,angle=-90]{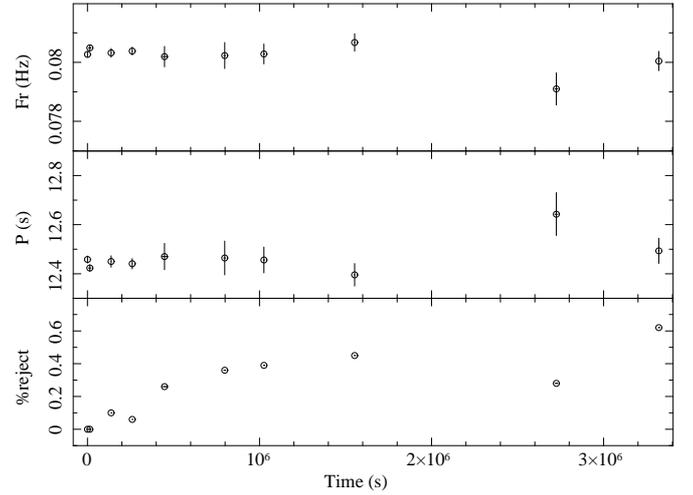}
\caption{Results of the search for pulsations in the \swift\,/XRT data. 
The upper (middle) panel show the determined frequency (period) of pulsations 
from the source in each observation as a function of time. In the bottom panel 
we show the percentage of the total number of simulated lightcurves 
(100) for each observation that did not result in a statistically 
significant ($\ge$3$\sigma$) detection of a spin period at the expected frequency. 
The reference time ($t=0$) is 55498.62225 MJD.}
\label{fig:pspin} 
\end{figure}

We used the $Z^2$-statistic technique \citep{buccheri83,mark02} to search for pulsations in all the 
\swift\,/XRT observations reported in Table~\ref{tab:swift}. In each of these observations   
the value of the source spin frequency was estimated accurately by fitting the peak in the 
$Z^2$-statistic periodogram with a model comprising a constant and a sync function. However, because the points  
in the $Z^2$-statistic periodogram are not independent, the error on the position of the centroid of the 
peak derived in this way could not be considered as a reliable estimate of the uncertainty 
on the source spin frequency. We thus simulated for each observation 100 event files 
with the same exposure and number of events as the real file (we assumed a constant flux 
during each observation). A sinusoidal signal  
at the frequency determined from the data was introduced 
artificially and then measured  {\it a posteriori} by using the $Z^2$-statistic  
(we investigated in each case a total of 1000 frequencies in the range 
0.078-0.082~Hz). The frequencies with a detection significance $\ge$3$\sigma$ determined with this 
method were then averaged to derive the value of the best spin frequency for each observation 
and the associated uncertainty at 1$\sigma$ c.l. (that we assumed equal to 1 standard deviation 
from the averaged value). We checked for each observation that the frequencies determined from the 
simulations with this method were all consistent within 1 standard deviation 
with the best value found by using the $Z^2$-statistic on the real event file. 
The results of this procedure are shown in Fig.~\ref{fig:pspin}. 
In the bottom panel of this figure, we also report for each observation the percentage 
of the 100 simulated lightcurves that did not result in a significant 
detection of a periodicity ($\le$3$\sigma$).  The observations characterized by a lower source 
flux or a shorter exposure gave, as expected, a higher percentage of failed detections. 
This suggests that the percentage reported in the figure can be considered 
a reliable estimate of the detection significance of the periodicity in each observation. 

The results reported in Fig.~\ref{fig:pspin} show that owing to the relative 
large uncertainties that affect most of the measurements, no significant 
variations of the source spin frequency could be measured  
during the \swift\ follow-up observations.  
This prevented any even tentative determination of an  
orbital solution for \igr\ through the  
pulse arrival time technique.   

\begin{figure}
\centering
\includegraphics[scale=0.37,angle=-90]{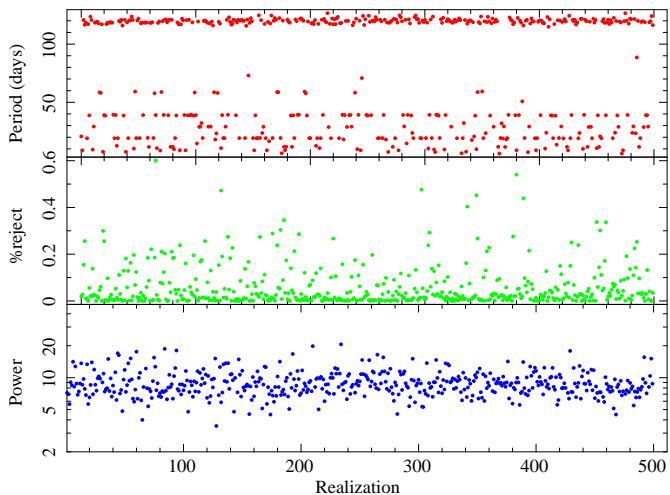}
\caption{Results of the simulations carried out to investigate 
the significance of the period at $\sim$120~days found from the \swift\,/BAT data. 
The upper panel shows the value of the period for each simulation (the number of 
the simulation is reported on the x-axis). The middle panel reports the probability 
that the detection of a period is false in each simulation (i.e., a value of 0.1 
corresponds to a detection significance of 90\% c.l.), and the lower panel the 
power of each peak in the corresponding LS diagram.} 
\label{fig:scargletotal} 
\end{figure}

To search for a confirmation of the suggested 
$\sim$117~d orbital period of \igr\ (see Sect.~\ref{sec:intro}),  
we therefore adopted an alternative method. We obtained the long-term 
lightcurve of the source from the \swift\,/BAT survey 
web-page\footnote{http://heasarc.gsfc.nasa.gov/docs/swift/results/bs58mon/.} 
binned snapshot-by-snapshot. 
All photon arrival times were corrected to the solar system barycenter 
with the task {\sc earth2sun}. 
Because the $Z^2$-statistic technique is particularly suited for un-binned 
event files rather than for lightcurves, we searched for periodicities in the 
BAT data by using the Lomb-Scargle (LS) periodogram \citep[we adopted the
fast algorithm described in][and scanned periods above 31~days with an 
oversampling factor of 8]{press89}. Because the uncertainties 
affecting the \swift\,/BAT data change significantly from pointing to pointing 
because of the different exposure time and off-axis position of the source with respect to the instrument 
aim-point, we also implemented the weighting scheme of the BAT lightcurve proposed 
by \citet[][in our case the correction factor $V_{\rm S}$ introduced to account for the 
source variability turned out to be negative and thus was set to zero]{corbet07}.  
The Lomb-Scargle periodogram was realized for both the original \swift\,/BAT lightcurve, 
and for the 500 lightcurves simulated from it. In performing the simulation, we assumed 
for each lightcurve a duration equal to that of the true \swift\,/BAT lightcurve and drew the source count-rate 
in each time bin from a normal distribution 
centered on the real measured count-rate\footnote{This assumption is justified for a coded mask 
instrument, because the source signal typically constitutes only a few percent of the large number of counts 
that characterize the background.}. We set the corresponding standard deviation equal to 
the measured error on the count-rate.  
The results of this analysis are reported in Fig.~\ref{fig:scargletotal}. 
The histogram of the periods determined from the simulated data showed a single 
prominent peak at $\sim$120~days, which thus turned out to be the most probable period 
(the simulation does not assume {\it a priori} a periodicity). 
However, we noticed that only 266 out of the 500 simulated lightcurves displayed a peak 
at a period of $\sim$120~days, and from these we estimated 
the best orbital period at $P_{\rm orb}$=120$\pm$2~days (68\% c.l.). 
Most of the remaining lightcurves (209) show a prominent peak at frequencies 
close to higher harmonics of the fundamental frequency. This distribution of the power 
among several harmonics would be expected according to the findings of 
\citet{corbet09}, who  showed that the folded lightcurve of the source 
was strongly not sinusoidal.  

\begin{figure}
\centering
\includegraphics[scale=0.6]{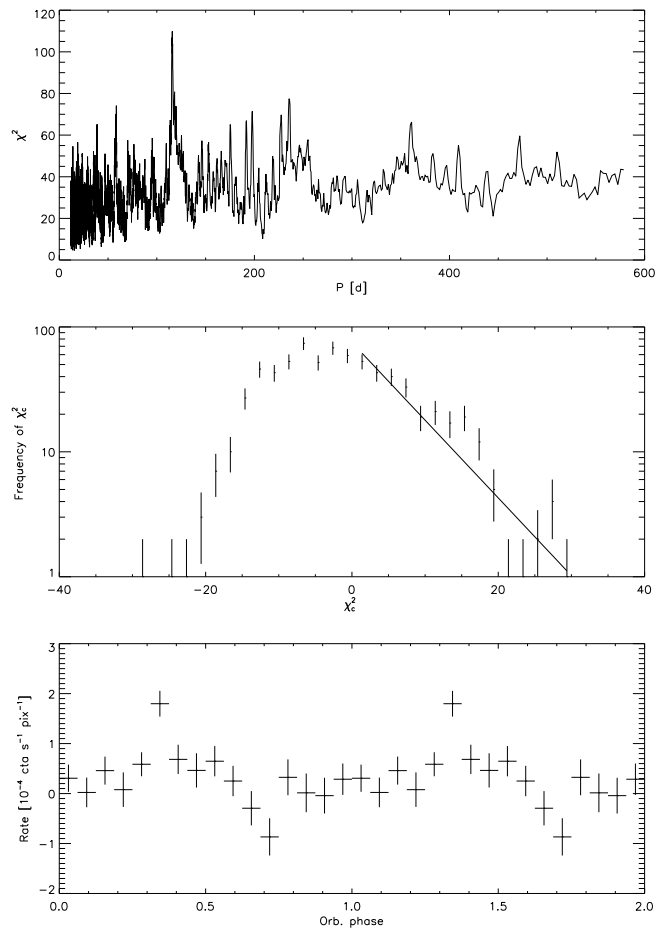}
\caption{Results of the search for periodicity in the \swift\,/BAT data obtained 
with the technique described by \citet{dai2011}.  
The upper panel shows the $\chi^2$ computed from the \swift\,/BAT lightcurve 
folded at different periods in the range 15-600~days. The most prominent peak 
is at $116.2\pm0.6$\,days. To estimate the significance of this peak, we followed 
the procedure described in \citet{dai2011}, and report the histogram of the detrended 
$\chi^2$ values in the middle panel. We show in the bottom panel the 
\swift\,/BAT lightcurve (15-50 keV) of the source 
folded at the best determined orbital period ($116.2\pm0.6$\,days). The reference time 
used is 54877.3957 MJD. Each phase bin in the plot corresponds to an integration time 
of three days.} 
\label{fig:scargle} 
\end{figure}

To refine our determination of the orbital period, we also looked for periodicities 
in the \swift\,/BAT lightcurve by exploiting the epoch-folding technique described by \citet{dai2011}. 
We folded the light-curve in 16 phase bins by using periods in the range 15-600\,d. The sampling 
in period was chosen to be 3 times finer than the resolution of the method, 
i.e. $\Delta\nu$=$P^2$/($N$$\Delta_\mathrm{BAT}$) \citep[here, $P$ is the period, $N$ is the number 
of phase bins, and $\Delta_\mathrm{BAT}$ is the 
data span length;][]{buccheri83}. We then computed for each of the folded lightcurves the 
corresponding $\chi^2$ and found the most prominent excess at a period of $116.2\pm0.6$\,days. The estimated 
chance probability that this detection would correspond to a spurious effect is 
about 1\% (see Fig.~\ref{fig:scargle}). 
In the bottom panel of Fig.~\ref{fig:scargle} we also show the \swift\,/BAT lightcurve folded 
at the best determined spin period. The reference epoch was chosen at 54877.3957 MJD, and we notice 
the presence of a prominent peak in the source count rate at phase 0.34$\pm$0.03. Considering an 
orbital period of $116.2\pm0.6$\,days, the epoch of the peak would agree with 
the prediction of \citet{corbet09}.  

We conclude that further observations are probably needed to firmly 
establish and eventually refine the orbital period of the source.  

\section{Discussion and conclusions}
\label{sec:discussion}

We reported on the 
\inte\ and \swift\ observations carried out for about two months 
after the newly discovered outburst of \igr\ 
on 2010 October 28 \citep{bozzo10}. 
During these observations, \igr\ displayed a smoothly decreasing 
X-ray flux with a decay time-scale similar to that observed during the 
outburst in 2009. 
Pulsations were clearly detected 
in the \inte\ data up to $\sim$50~keV and in the \swift\,/XRT data. 
Owing to the relatively low X-ray flux of the source 
during most of the XRT observations, the accuracy with which we 
could measure its spin period was far too low to determine an orbital 
solution for \igr\ with the pulse arrival times technique. 
Instead, we found a marginal detection of this periodicity with  
both the LS periodogram and the epoch-folding technique. 
We found the best orbital period at 116.2$\pm$0.6~days, with 
a chance probability of 1\% to be a spurious detection. 

As \citet{rodriguez09b} remarked, the spin and orbital 
period of \igr\ (if confirmed) would nicely place the source in the 
part of the $P_{\rm spin}$-$P_{\rm orb}$ diagram \citep{corbet86} populated by the so-called 
Be X-ray binaries. The \inte\ and \swift\ observations of the source carried out during 
the outburst in 2010 provide further support to this hypothesis.  
The X-ray pulsations at $\sim$12.5~s and the high rms fractional amplitude 
measured up to 50~keV are typical of Be X-ray binaries 
and are not usually observed in the SFXTs \citep[the $\sim$6~days long outburst 
shown in Fig.~\ref{fig:lightcurvestot} would also be typical for a Be X-ray binary; see e.g., 
][for recent reviews]{ziol02,reig11}.  
Among all the candidate and confirmed sources in the SFXT class ($\sim$15 objects), 
so far only\footnote{Note that Bozzo et al. (2011, submitted) showed that the NS hosted in the 
SFXT IGR\,J18410-0535 might not be pulsating at $\sim$4.7~s as it was previously reported.} 
the intermediate SFXT IGR\,J18483-0311 showed pulsations with a period similar 
to that of \igr\ \citep[$\sim$21~s detected up to $\sim$40~keV,][]{sguera07}.  
However, the smooth and gradual decline of the X-ray flux from \igr\ during two months 
spanned by the \swift\ observations seems to be different from the behavior displayed 
by the SFXT sources. Indeed, these sources  
usually show a rapid (thousands of seconds) decline of the X-ray flux of 3-5 orders of magnitude  
after the occurrence of an outburst. In those cases in 
which a detailed monitoring could be carried out with \swift\,/XRT after the occurrence of a 
bright event, it was also noticed that a pronounced variability in the X-ray flux (a factor $\sim$100), 
accompanied by significant changes in the spectral parameters (e.g., the absorption column density),  
characterize the source emission during the entire orbit \citep[see e.g.,][]{bozzo08,bozzo09,sidoli09}. 
The slow decline of the X-ray flux from \igr,\ combined with the fairly small changes in the spectral parameters 
across the entire \swift\ monitoring, suggest that the behavior of \igr\ is closer to that of the 
Be X-ray binaries.  
We thus conclude that \igr\ is most likely a Be X-ray binary and not an SFXT.

\begin{acknowledgements} 
We thank the \swift\ team for the 
prompt scheduling of all the follow-up observations 
of \igr,\ and an anonymous referee for useful comments. 
This research has made use of the XRT Data 
Analysis Software (XRTDAS) developed under the responsibility 
of the ASI Science Data Center (ASDC), Italy. 
\end{acknowledgements}

\bibliographystyle{aa}
\bibliography{igrj19294}

\end{document}